\documentclass[aps,prl,twocolumn,showpacs,superscriptaddress,groupedaddress]{revtex4}  % for review and submission
\usepackage{graphicx}  % needed for figures
\usepackage{dcolumn}   % needed for some tables
\usepackage{bm}        % for math
\usepackage{amssymb,amsmath}   % for math
\usepackage{array}
\usepackage{units}
\usepackage{color}
\usepackage{natbib}	

\definecolor{darkgrey}{RGB}{50,50,50}
\graphicspath{{}{.}}
\begin{document}

\title{Exponentially fast Thinning of Nanoscale Films by Turbulent Mixing}
%\input author_list.tex       % D0 authors (remove the first 3 lines
                             % of this file prior to submission, they
                             % contain a time stamp for the authorlist)
                             % (includes institutions and visitors)
\author{M. Winkler}\affiliation{University of Potsdam, Potsdam, Germany}
\author{G. Kofod}\affiliation{University of Potsdam, Potsdam, Germany}
\author{R. Krastev}\affiliation{NMI Uni Tübingen, Reutlingen, Germany}
\author{S. St\"ockle}\affiliation{Max--Planck Institute for Colloids and Interfaces, Potsdam, Germany}
\author{M. Abel}\affiliation{LEMTA - UMR 7563 (CNRS-INPL-UHP), 54504 VANDOEUVRE, France}

\date{\today}

\begin{abstract}
Films are nanoscopic elements of foams,
emulsions and suspensions, and form a paradigm for nanochannel transport that
eventually tests the limits of hydrodynamic descriptions.
Here, we study the collapse of a freestanding film to its equilibrium.
The generation of nanoscale films usually is a slow linear process;
using thermal forcing we find unprecedented dynamics with
exponentially fast % accelerated ??
thinning. The complex interplay of
thermal convection, interface and gravitational forces yields optimal
turbulent mixing and transport.
Domains of collapsed film are generated, elongated and convected
in a beautiful display of chaotic mixing.
With a timescale analysis we identify mixing
as the dominant dynamical process responsible for exponential thinning.

\end{abstract}

\pacs{47.61.-k,47.57.Bc,82.70.Rr, 47.51.+a}
\maketitle

Thin film dynamics is governed by \textit{gravitational},
\textit{capillary} or \textit{interfacial} forces,
inducing the \textit{disjoining pressure}.
The latter combines long- and short-range molecular forces:
electrostatic, Van der Waals (VdW), and steric forces \cite{mysels1959soap,exerowa1998foam,oron1997long}
and strongly depends on the distance between the interacting surfaces.
Whereas films on substrates are established in industry and research,
freestanding thin liquid films (foam films) are
important in nature and technology, but still
provide a challenge in experiments and theory alike \cite{reiter1998artistic}.
Consequently, the study of foam films is of central interest
\cite{exerowa1998foam,kellay1995experiments}, in particular their formation.
The latter is primarily governed by the \emph{thinning behavior}.
We present a novel approach to thinning
of \textit{vertically oriented, freestanding,  nonequilibrium, thermally forced}
foam films (i.e. driven by a temperature gradient) \cite{bruinsma1995theory}.
By turbulent convection \cite{Ahlers-09}, we mix the fluid,
i.e. we stretch and fold material lines \cite{Ottino} . This leads to entirely new dynamical
behaviour as we will demonstrate below.

In the following we confront foam films \textit{with} and \textit{without} forcing.
\textit{Without perturbing forces}, two stable equilibria may occur,
depending on the bulk solutions' chemistry and surface active agents (surfactants):
Common Black Films with a thickness of more than \unit[10]{nm} are formed
when electrostatic interactions
balance the dominant Van der Waals force, gravity, and capillarity \cite{exerowa1998foam};
Newton Black Films are stable  with  a thickness of less than
\unit[10]{nm}, due to repulsive short range steric forces
\cite{exerowa1998foam}. We refer to both
as Black Film (BF).

Since the BF is the stable phase, its progression
into the thick film can be considered as a front propagation problem \cite{van2003front},
with BF front velocity, or thinning speed, $v_{BF}$; in contrast to other fronts (e.g. chemical, or biological),
fluid is conserved: the front is limited by the bulk  draining velocity of the
film lying below.
Therefore, the deposition of fluid in the thick bottom meniscus and the subsequent outflow dominate the thinning speed.

\begin{figure}
\includegraphics[width=\columnwidth, draft=false]{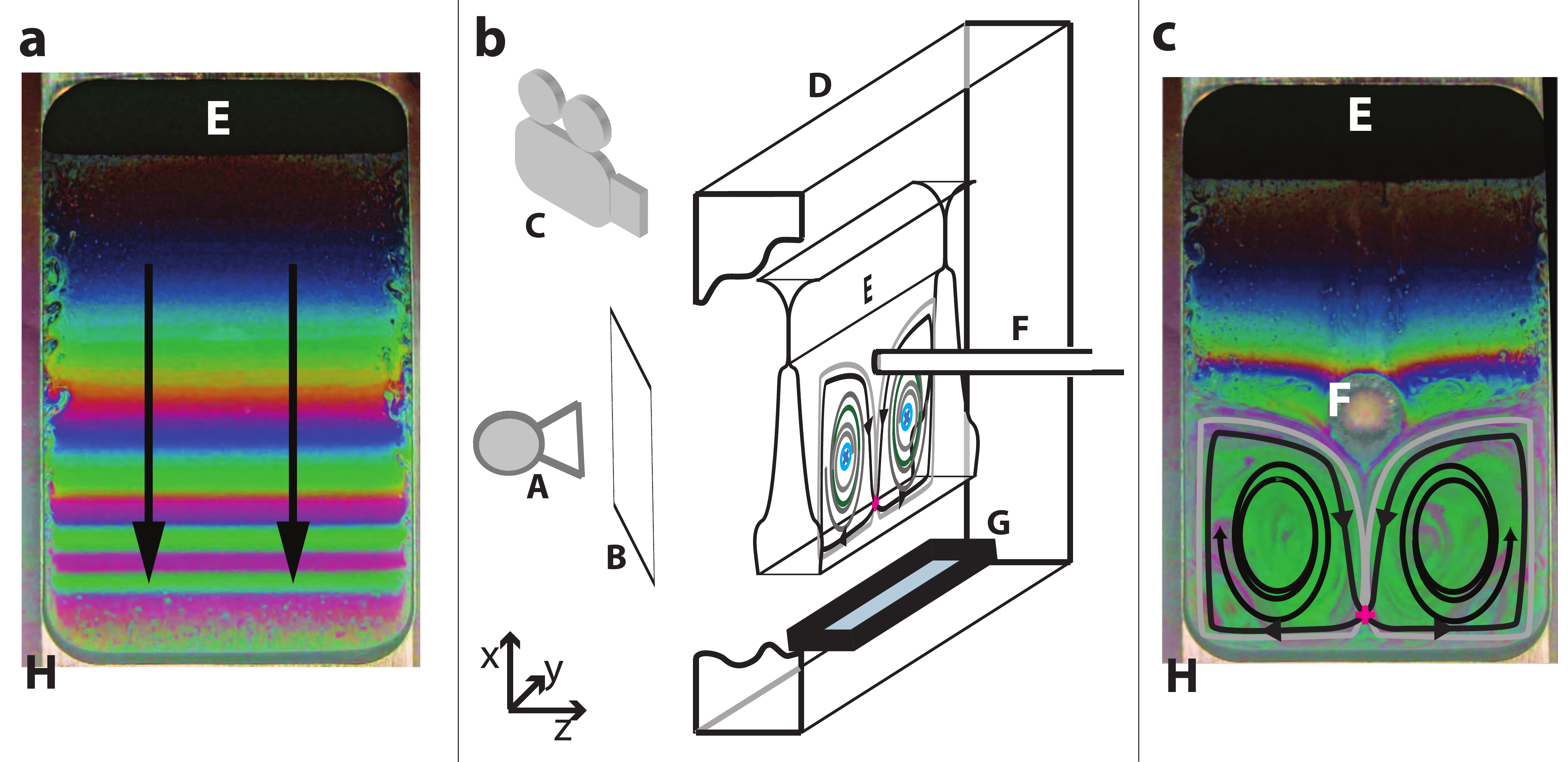}
\caption{
The (left and right) snapshots show the reflection of a diffuse light source off the foam film surface.
The color, as seen by the naked eye, repeat from top to bottom
with the increasing film thickness.
\textbf{a} Foam film in aluminium frame - thinning without thermal forcing.
\textbf{b} Perspective view .
\emph{A:} Light source, \emph{B: }diffusor screen, \emph{C: }video camera,
\emph{D: }casing to prevent evaporation, with plexiglass window,
\emph{E: } BF area
\emph{F:} Cooling copper needle at $-169\,^\circ$C,
% entering through a fitting hole in the back plate,
\emph{G: } Soap reservoir,
\emph{H:} Aluminium frame %for foam film support.
The convective pattern, established with thermal forcing is sketched,
along with the BF and the thick, wedge-like region. The latter decreases in the course of time.
\textbf{c} Foam film in aluminium frame - thinning with thermal forcing.
The convection rolls are in the lower half of the frame, above the cold needle a stably stratified region has formed.}
\label{fig:setup}
\end{figure}

This in turn is effected by the foam film profile:
it consists of a large, flat bulk region and a meniscus, the Plateau border,
which is much thicker, cf. Fig.~\ref{fig:setup}, a.
Whereas the drainage in the bulk region follows a slow Poiseuille flow \cite{stoyanov1997motion},
the main transport in the Plateau border \cite{saint2004quantitative},
is well-known as marginal regeneration and shows a rich variety of
patterns, cf. \cite{supplement,nierstrasz1998marginal}. Eventually, the whole film is thermodynamically stable as a BF.
A slow drainage, steady-state profile and
thermal effects have been predicted earlier, in experiments with
evaporation and comparatively uncontrolled chemistry
\cite{mysels1959soap}.

Here, a foam film is brought far into nonequilibrium by
pointlike thermal forcing, cf.~Fig.~\ref{fig:setup}, b. The additional force
counteracts gravitation and capillarity,
and changes the dynamics from linear to
exponential.
After a transient period, a global convective flow with complex
dynamic behavior is established, marked by two-dimensional turbulent mixing combined with surface
instabilities. Image recording allows for a characterization of the flow field,
global mixing properties,  and eventually the thinning law,
as the film progresses towards a global BF state.
As it turns out, the observed behavior is generated by the unique combination of dynamical processes within the nanoscale film.

\paragraph{Experiment}

The experimental setup consists of a vertical rectangular aluminium frame
with rounded corners, \unit $45 \times 20${ mm},
enclosed by an atmosphere-preserving cell with a glass window for video recording,
cf.~Fig.~\ref{fig:setup}, for more details and video, see \cite{supplement}.
Both Common and Newton BF are optically transparent, and show  a subtle
difference in reflectivity which is not resolved by our setup.
Thermal forcing is effected by inserting a cooled copper needle (radius \unit[1]{mm})
at the film center (\unit[$T=-169$]{$^\circ$C}),
The needle enters the cell through a fitting hole.
Ambient temperature was constant at \unit[20]{$^\circ$C},
with  a Rayleigh number $Ra\sim 10^6$, typical for turbulence.
The temperature across the film (z-direction, cf. Fig.~\ref{fig:setup}) is approximately constant,
and the Marangoni number $Ma\simeq 0$ \cite{supplement}.

The solution from which the liquid film was drawn consists of
the surfactant n-dodecyl-$\beta$-maltoside ($\beta\mathrm{-}C_{12}G_2$),
prepared with filtered deionized water and stabilized with 25~$\%_{\rm vol}$ glycerin
\cite{stoeckle2010dynamics}.
The entire film area is observed optically by a conventional video camera \cite{supplement}.

Our Vertically oriented foam film is produced initially thick (500 -- 5000 nm)
by pulling it with a glass rod from the reservoir, cf.~\cite{supplement}.
Quickly, a wedge-like profile develops with BF in a small
region, with a sharp horizontal boundary towards the thick film below,
cf. Fig.~\ref{fig:setup}, 'E'.
% The description of the time evolution
% of this BF region.
The interference of incident and reflected light yields
a striped pattern, which can be used to infer the film thickness.
Each color cycle corresponds to the multiples $n$ of the smallest negative interference condition
$(2\cdot n+1) \lambda \eta = 4 h \cos \Theta$, where the refraction index, $\eta$, is
assumed to be temperature-independent; $\Theta$ is the angle of incidence \cite{atkins2010investigating}.
The velocity is measured by  color imaging velocimetry (CIV) \cite{Winkler-Abel-2012} with $u \sim  0.02\,\unitfrac{m}{s}$.
% The main concern is the thinning speed, so extra care is spent
% on its exact determination:
%For maximum accuracy, color tables and contrast of
%the video data was processed to pixel-level accuracy.
% of  the position of the interface between BF and thick film.

\paragraph{Results}

The experiments described were repeated several times in order to
check reproducibility. The main sources of variation are i) initial conditions,
%this has been counteracted by starting comparison when the first BF had formed,
ii) chemistry, which had been controlled by the highest accuracy known to us \cite{stoeckle2010dynamics},
iii) the probabilistic character of the flow, this is a intrinsic property and
can only be compensated by statistical analysis.

\begin{figure*}
  \includegraphics[width=0.8\textwidth, draft=false]{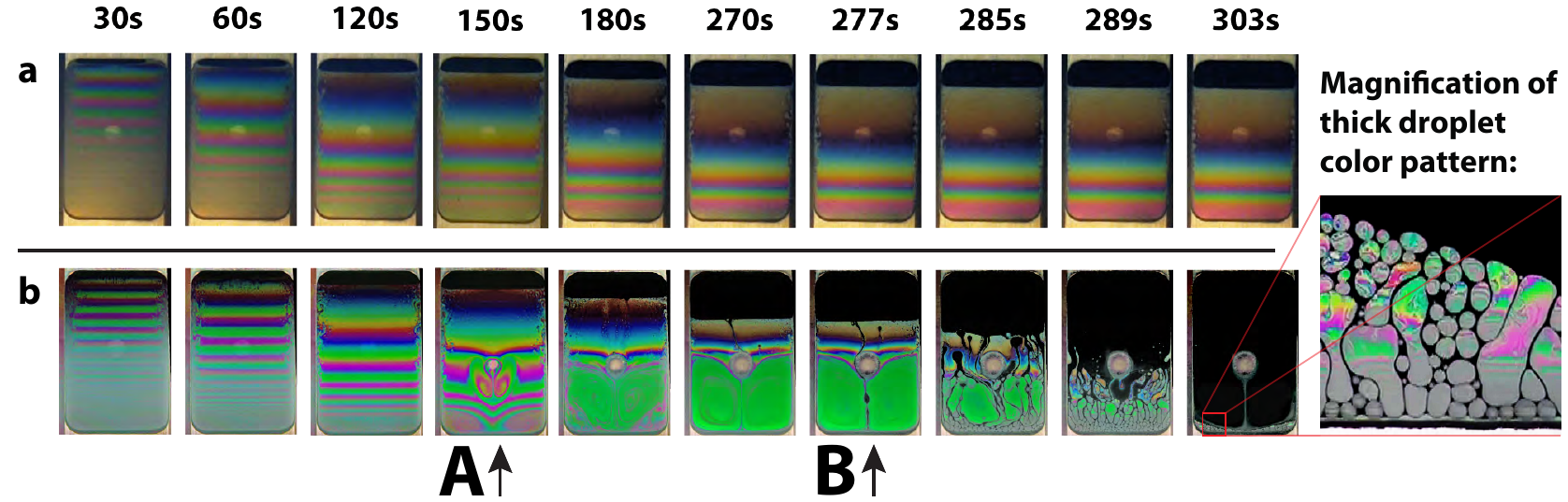}
\caption{ \textbf{Image sequences of foam film thinning.}
 Thinning behaviour of a foam film \textbf{a} \textit{without}, and \textbf{b} \textit{with} thermal forcing.
While the undisturbed thinning (top) evolves slowly, the thermally driven flow (bottom)
exhibits convection, with rapid transport and mixing.
\textbf{A:} start of convection, \textbf{B:} onset of exponential thinning.
One clearly recognizes the black film spots which
are stretched and folded, such that a rapid conversion the film evolves to the
BF equilibrium phase. The magnification at the right shows the
following:
domains of thick film are surrounded by black film, such that the combination of
forces produces an approximately spherical cap, with the lower,
plane part parallel to the foam film. Consequently, the
$(2n+1)\frac{\lambda}{4}$-condition does not hold anymore since it is based on
the reflection between two parallel planes.
}
\label{fig:thinningsequence}
\end{figure*}

\textit{Without} thermal forcing,
the initially thick film (micron scale) is drained by
gravitation and capillarity, thereby evolving towards its equilibrium BF thickness.
This thinning is limited by Poiseuille flow \cite{couder1989hydrodynamics,bruinsma1995theory} with the velocity $v\simeq \frac{g\,h^2}{2\nu}$.
In the bulk, $h\sim\,$\unit[$10^{-6}-10^{-8}$]{m}, and
$v\sim 10^{-6}-10^{-10}\,\unitfrac{m}{s}$.
On the vertical boundaries, the so-called  Plateau border with a
meniscus of $h\sim 10^{-5}-10^{-6}\,\unit{m}$ allows for a faster transport
at $v\simeq 10^{-4}-10^{-6}\,\unitfrac{m}{s}$, it is the dominant mechanism
for our aspect ratio. The film in this frame reaches equilibrium, i.e. complete BF phase after approximately ~\unit[5]{h}.
A corresponding series of snapshots is given in Fig.\ref{fig:thinningsequence}, a;
the dominant transport can be observed by a
Kelvin-Helmholtz-like instability at the Plateau border.

We clearly observe a linear behavior with constant velocity
$v_{regular}\simeq$\unitfrac[18]{$\mu$m}{s},
in accordance with \cite{stoyanov1997motion}.
The very first quantitative  thinning experiments \cite{mysels1959soap}
reported an exponential dynamics on the time scale of $\tau = 1000\,\unit{s}$,
in order to check this, we measured the thinning with open atmosphere:
for our chemistry we find a slow, exponential thinning ($\tau = 4000\,\unit{s}$), which we
attribute to evaporation; chemistry of the surfactants plays a role, too
cf.\cite{mysels1959soap,supplement}.

\begin{figure}
\includegraphics[width=\columnwidth, draft=false]{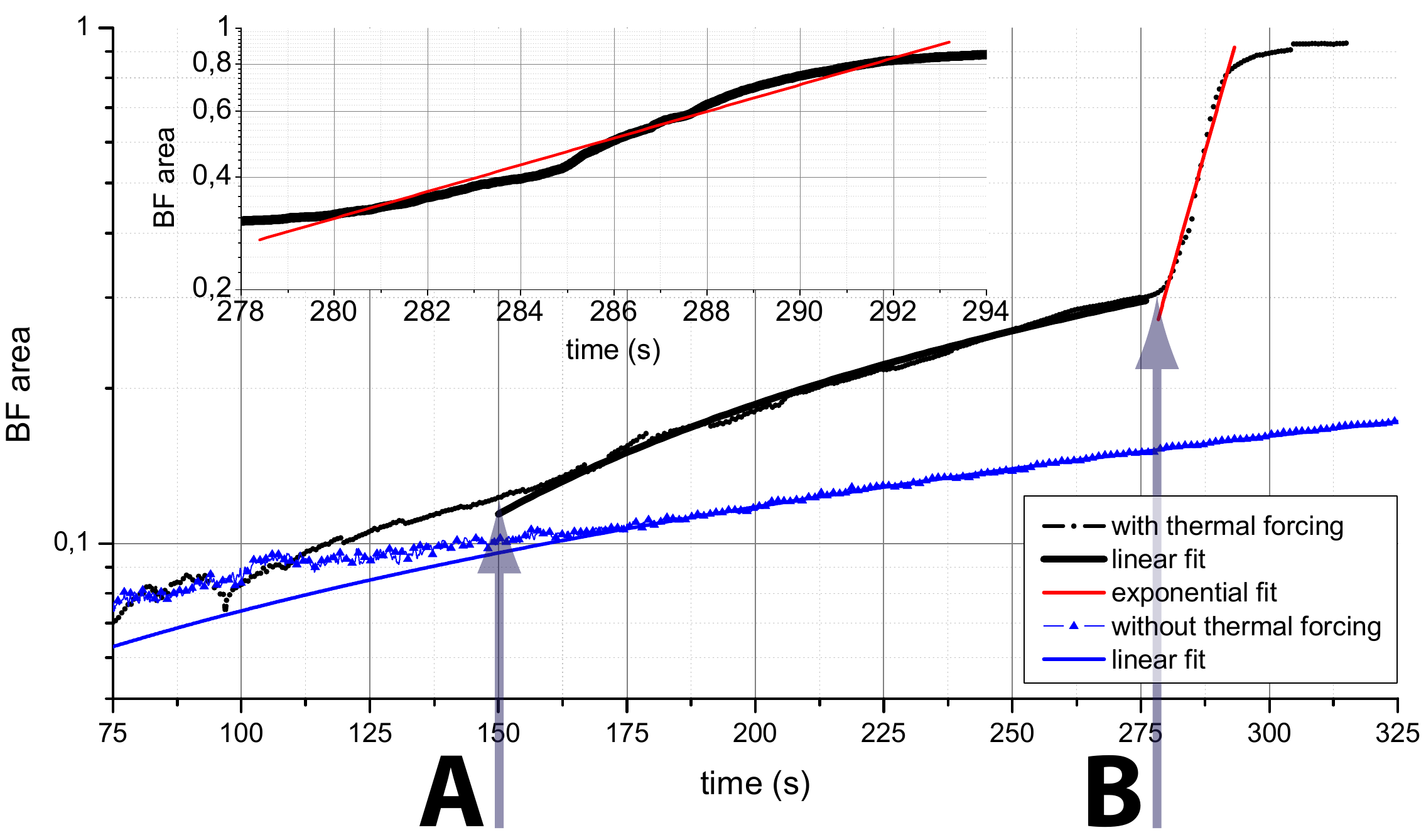}
\caption{\textbf{Temporal evolution of BF area: linear versus exponential thinning.}
%Regular thinning (triangles) versus thinning with \TF{thermal forcing} (circles), and the lines are linear fits.
\textcolor{blue}{\textbf{blue triangles}}: \emph{without} thermal forcing, the thinning is perfectly linear in time.
\textcolor{darkgrey}{\textbf{ black circles}}: \emph{with} thermal forcing, one observes three
different states: for times $t<150s$ (transition at \textbf{A}), no thermal forcing is applied and
the convection has not yet started. For times $150\,s<t<280\,s$, the convection is established, but
BF spots are not stretched into filaments, consequently thinning is linear but faster.
Finally, at $t=280\,s$ the transition
to exponentially fast thinning can be observed, marked by \textbf{B}. BF spots and turbulent mixing
are acting inside the convection zone. Solid lines are linear fits.
The inset demonstrates the exponential behavior by a high-quality fit for $1-A = exp(-t/\tau)$,
for this experiment with $\tau=$\unit[4.76]{s}.
}
\label{fig:exponentialfit}
\end{figure}

%thermal driving
\textit{With} thermal forcing (\textbf{(A)} in Fig.~\ref{fig:thinningsequence} and Fig.~\ref{fig:exponentialfit}), %, caused by the cooled copper needle,
the film is cooled down rapidly in the center and a small,
frozen region appears surrounding the needle. Thermal energy is
exchanged at its boundaries and heavy convection starts. A weakly turbulent
flow is rapidly established with 2 symmetric, dominant eddies,
cf.~Fig.~\ref{fig:setup}, c. As a result of this convection, two different
regimes are observed: faster, but still linear, and exponential thinning.
We explain here the dynamics of \emph{one} data set, presented in Fig.~\ref{fig:thinningsequence},
in total we have analyzed 15 runs, all with the same reproducible dynamical behaviour
\cite{supplement}. Our result is a mean time constant with corresponding deviation.

In the linear regime (\unit[150]{s}--\unit[277]{s}) the velocity of the BF front is still constant,
however the magnitude is increased by a factor of 3 to $v'_{BF}=$\unitfrac[66]{$\mu$m}{s}
(cf.~Fig.~\ref{fig:exponentialfit}, black line).
The enhancement is explained as follows:
Fluid is transported by convection vertically from top to bottom, however the
outflow into the reservoir is slower than the deposition rate of fluid by the convection.
Necessarily, a vertical upflow at the left and right boundary is generated, causing suction of fluid from the bottom.
This results in a greater, almost uniform thickness of  the
 convection area. This  increases the downward Poiseuille flow in the bulk film, such that the
 BF front velocity $v_{BF}$ increases.

 The second, exponential regime starts at $t\simeq$ \unit[277]{s}
(\textbf{B} in Fig.~\ref{fig:thinningsequence}), its dynamics is explained as follows:
The BF spots arise spontaneously due to thermal and mechanical fluctuations,
and are subsequently transported around by the convection.
A BF spot is basically a domain of BF (5--20 nm)
surrounded by much thicker film (100--1000 nm).
This BF spot can only grow by transporting liquid away from the spot
through the surrounding film. This transport happens again
under Poiseuille flow conditions, due to the nanoscale
thickness of the film.

To produce a local BF spot, mechanical and/or thermal fluctuations
must exceed the energy needed to collapse. Here, we can only give an
estimate, following \cite{diamant2010localized}: the pressure balance
on the line between BF spot and surrounding thick fluid is given by
$ p= -\sigma \nabla^2 h -\Pi(h)$ with $\sigma$ the surface tension.
We estimate $\sigma \cdot \nabla ^2 h \simeq$ \unit[3.4]{kPa}, and
measured $\Pi(h)$ to be \unit[10]{kPa} \cite{supplement}.
The pressure itself cannot be calculated analytically, however, we
can estimate mechanical stresses and thermal energy per front volume as
$\tau_{mech} \simeq$\unit[0.6]{Pa} and $\Delta p_{thermal}\simeq $\unit[6.3]{kPa},
half the pressure to overcome, \unit[3.4+10=13.4]{kPa} \cite{supplement}. Fluctuations then
can sometimes exceed the threshold, and the relatively seldom occurrence
of spontaneous BF formation supports our estimation.

A single BF spot is lifted up like a bubble due to gravity, and
velocity gradients acting on its line interface
give rise to elongation and contraction, at rate \unit[0.3]{$s^{-1}$} and \unit[2.52]{$s^{-1}$}, respectively.
This \textit{stretching} leads to the formation of thin BF filaments beyond a certain threshold of size and speed
(cf. Fig.~\ref{fig:thinningsequence} approx. 285 s, and
 \cite{supplement}) .
These filaments are then \textit{folded} following the turbulent convection.
Consequently the two basic ingredients of mixing are present,
and the typical exponential separation of trajectories is reflected in the thinning law.
As a result BF spots are lifted to the
top to join the BF area, droplets descend due to gravity and eventually flow out into the bottom reservoir.

Now, we compare the timescales for thinning with and without thermal forcing.
Essentially, the thickness varies due to i) density variation according to the local
temperature; this effect is of the order
$\frac{\Delta \rho}{\Delta T \cdot \alpha} \simeq 0.005 $
ii) advection of $h$ with a ratio $5/500$ (final vs.~initial thickness).

In the following we relate our observation to the governing dynamical equations
in order to find a relation between thinning rate and convective mixing.
The thinning process obeys
an exponential decrease of the transient, thick phase:
$ \dot{A} = \gamma (A_0-A)$,
with $A$ the area covered by thick film (Fig.~\ref{fig:exponentialfit}). We find the
thinning speed $v\simeq \dot{A}/L_y$ with $L_y$ the lateral extension, and 
a thinning timescale, $\tau=1/\gamma = \unit[6.19]\pm\unit[2.63]{s}$  -
this is 650 times faster than usual.
We assumed a boundary approximately normal to gravity and space directions
as indicated in Fig.~\ref{fig:setup}, b.

% In general, the timescale depends on geometry and flow
% properties,  $\tau = \tau(v,h,L_x,L_y)$.
On the other hand, the change of BF area is reflected in the
change of local ``control'' volumes $dV = h(x,y) dA$: since the total area $A_0=L_x\cdot L_y$
is constant, the only volume  change  is due to thickness evolution,
such that  the BF front can be obtained using
$ V=   \int_{A_0} dxdy\; h(x,y)$.
We are interested in the transient regime, and thus can use
the evolution equation for $h$ given in \cite{oron1997long}:
%with a disjoining pressure which does not yield a
%stable BF, but describes correctly the transport of $h$:
\begin{eqnarray}
 \dot{V} &=& \int_{A_0} dxdy\; \dot{h} = -\int_{A_0} dxdy\; h(x,y)\nabla_2 u(x,y)
\end{eqnarray}
with $\nabla_2$ the two dimensional divergence and $u$ the corresponding velocity field.
Note that i) $\dot{h}$ is the convective time derivative, i.e.~the local growth and
ii) the flow is highly compressible in 2D, justifying the above equation.
The average thickness, denoted by brackets, evolves as
$
 \dot{h} = -h \left <\nabla_2 u \right >
$
\cite{Erneux-Davis-93}.
Identifying $\tau$ with $1/{\left <\nabla_2 u \right >}$ we have related
the mean convection with the thinning law.
Further, we can relate this result to mixing theory \cite{Ottino}, where
the (eulerian) exponential separation of fluid points is contained
in the mixing efficiency, defined as the mean stretching rate, normalized by
the stretching tensor $1/2(\nabla u + (\nabla u)^T)$. Obviously, the
thinning time is directly proportional to this quantity, and
consequently to the mixing properties of the flow \cite{Winkler-Abel-2012}.
The quantitative estimate of the mixing efficiency requires
more data than available, however we can test for consistency
of the sign.
% $e_{l} = \lim_{\Delta x_0 \to 0} \frac{\Delta x_{t}}{\Delta x_0}$
% The flow is dominated by two local structures, the BF spots and
% the BF filaments.
The BF spots and filaments, respectively, expand and thus $\left <\nabla_2 u\right> > 0$,
i.e.~the thickness $h$ decreases exponentially.
% In the absence of a restoring force, this would lead to rupture;
% in our case the divergence of the velocity
% converges to zero for the thickness going to its equilibrium BF value.
%due to disjoining pressure.
% Due to local mass conservation, if the thickness decreases in one area,
% there must be a corresponding increase in
% nearby regions, leading to ``rims'', similar to the ones found in evaporative dewetting \cite{diamant2010localized}.
% They are also subject to gravitation, and fluid flows to the bottom. The time scale
% is now set by convection in contrast to Poiseuille flow.

We can test our approach by calculating the parameters and time scales of the involved processes:
thermal forcing, gravitation, steric and electrical forces (including Van der Waals), capillarity, mechanical strain (dissipation).
To estimate the important effects, we calculate Rayleigh-, Capillary-, Marangoni-, Biot-, and Prandtl
numbers ($Ra$, $Ca$, $Ma$, $B$, $Pr$), and the dimensionless Hamaker constant $\tilde{C_H}$.
This leads to the following values: $Ra\sim 10^6$, $Ma\sim 0$, $Ca \sim 10^{-3}$,
$B\sim 10^{-4}$, $Pr\sim 17.5$, $\tilde{C_H}\sim 10^{-4}$, cf. \cite{supplement},
% $Ra = g \beta \Delta T L^3/\nu/\kappa_{th}\sim 10^7$,
% $Pr\sim 7$,
% $Ma=\partial \sigma/\partial T  \Delta T H/\mu/\kappa_{th} \sim 10^3$, $B\sim $, ($\partial \sigma/\partial T \sim 1$)
% $B= \alpha H/\kappa
from which we deduce that the dominant force is due to temperature.
% $ \tau_{therm} \sim \frac{\rho c L^2 }{\Delta T\kappa_{th}} \sim  \todo{??? s}$,
% $ \tau_{grav} \sim \frac{l^2 \mu}{H^2\rho g L}\sim   10^{-2}s$,
% $ \tau_{disj} \sim \frac{\mu\,l^2}{A}\sim  10^{-2}s$,
% $ \tau_{cap} \sim \frac{1}{l}\frac{\mu\,l^3}{H^4 |\sigma|}	\sim 10^{-3} s$.
The forcing produces convection, with two typical time scales: a global one, the characteristic turnover time
$ \tau_{conv} \sim \frac{2L}{u} \sim 5\,s$ and a local one, the time scale
of the strain, $\tau_{strain}\sim 0.4\,s$.
% , responsible for the local generation of BF spots.
The latter is smaller than the first, consequently the large-scale motion
sets the time constant for exponential decrease of thick film area.
We measure $\tau= \unit[6.19]\pm\unit[2.63]{s}$, which  coincides very well with the convective scale,
given the roughness of our assumptions and the complexity of the process.

% \todo{this is based on $\partial_z p = \frac{\Delta h \rho g} = 10^{-2} m \cdot 10^3 kg/m^3\cdot 10 m/s^2 =100 Pa/m$
% and $v=h^2 \partial_z p / \mu$ with $h\simeq 10^{-6} m$}

% One last question remains, regarding the generation of the BF spots and
% the complex folding of the channels.

\paragraph{Conclusion}
We have demonstrated the dramatic acceleration of the thinning of a foam
film by thermal forcing.
We use novel experimental techniques to control the
properties of the thin liquid film in a precise manner, i.e. a measurement
chamber that inhibits evaporation from the film surface.
The analysis of the fluid flow is local and global,
by tracking BF spots to measure \emph{local} stretching, folding and \emph{global} thinning rate.

The thermal driving leads to convection, which changes the dynamics of thinning
from linear to exponential. This is not only substantially faster,
but qualitatively different from
the classical, thermally homogeneous, thinning process.
This result can be explained: the growth of
BF area is driven by the pressure gradient at the interface with the
non-equilibrium phase. Thus the interface length determines the overall thinning
rate. By advection and stretching of BF spots, trails are left behind
which increase multiplicatively the interface length resulting in the
exponential thinning behaviour.
Efficient transport and filamentation are guaranteed by the mixing properties of
the flow.

We plan to widen our work to a Rayleigh-B\'enard setup where the influence of interfacial
forces can be studied, or, more technically, in a controlled mixing of substances in a quasi 2D
setup. Further, with chemical control one can tune the film thickness and thereby observe the dependence
of thermodynamical and hydrodynamical properties on the thickness, eventually testing
the limits of continuum description. A comparison with quantum calculations will be
a formidable task for future research.

We acknowledge initial stimulus by Nicola Abel and discussions with L. Biferale,
D. Lohse, M. Sbragaglia and K.Q. Xia.
This research was supported in part by the NSF under Grant No. NSF PHY05-51164
and by cost action 806 ``particles in turbulence''. GK thanks the German Federal
Ministry of Education and Research (BMBF) for support via grant No. 03X5511
KompAkt (WING-NanoFutur).

\end{document}